\documentclass[preprint,12pt]{elsarticle}

\usepackage{amssymb}

\journal{Advances in Quantum Chemistry}

\begin{document}

\begin{frontmatter}



\title{Quantum Monte Carlo with ground-state input to investigate methane reactions on supported metal film catalysts.}


\author{Philip E. Hoggan}

\address{Institut Pascal, UMR 6602 CNRS, BP 80026, 63178 Aubiere Cedex, France.}

\begin{abstract}
\hskip4mm Nowadays, there is pressing demand for sustainable energy sources, or clean and 'green' fuel and hydrogen is a perfect candidate. It can be made by dissociating methane with the energy input compensated by metal-hydrogen bond formation. Industry uses Nickel supported on $\gamma$-alumina for this rate-limiting step.

This work shows how to use Quantum Monte Carlo calculations to take (user-guided) information summarising chemistry of heterogeneous catalysis from a generic Jastrow factor, that allows electron correlation to be included in the density. This, in turn, gives accurate values of properties, in particular the activation barrier for stretching one methane C-H bond, before adsorbing CH$_3$ and H on vicinal nickel atoms in the close-packed (111) plane.

The CH$_3$ co-ordinates to a hollow site defined by three Ni atoms and, new to this work, another Ni-atom co-ordinates the H-atom from the dissociation.
\vskip14mm

\end{abstract}
\vskip10mm

\end{frontmatter}


\newpage
\section{Introduction}

Recently, doubling the capacity of Paris Charles de Gaulle (CDG) airport has been postponed until hydrogen-powered planes become available for routine service to compensate environmental impact. Now is the time to develop mild conditions for selective hydrogen synthesis.

One efficient route to hydrogen is noble-metal surface catalysis of the methane pyrolysis.

\vskip4mm
CH$_4$ + Ni(111) $\rightarrow $  CH$_3$$ -Ni$ + $ H-Ni$
\vskip4mm

We have previously studied the first (water on CO attack) reaction step involved, on Pt(111) by Quantum Monte Carlo (QMC) methods. This requires location of the Transition-State (TS) for water addition to pre-adsorbed CO.

Since bond dissociation and formation are both involved in this crucial step, it was previously studied by us by quite time-consuming QMC, using high-level Multi-Reference Configuration Interaction (MRCI) wave-functions as input for a molecular active site embedded in the periodic metal-catalyst lattice \cite{hjcp,shar}. In this study, our objective is to ascertain how well this wave-function quality can be approached using a ground-state product with a sophisticated generic Jastrow factor, established by using chemical insight.

The key to this is suitably correlating the QMC trial wave-function with a generic Jastrow factor. Since it is time-consuming to optimise the full Jastrow (with terms specific to each atom) and even more so to propagate the ground-state by Diffusion Monte Carlo (DMC), we describe a tailor-made optimal single-determinant wave-function with reduced parameter-set generic Jastrow (SD-RPS-gJ), in particular focussing on the CO carbon to account both for adsorption and its role as the site of nucleophilic addition but including each molecular atom individually, whilst grouping those of our catalyst lattice and its surface. Three-body (electron pair+nucleus) terms are shown to be by far the most sensitive of SD-RPS-gJ coefficients to bond-breaking and formation. This simplified wave-function is then bench-marked in comparison to previous fully-optimised results.

Lately, error cancelation in such QMC work involving the same atoms in TS and asymptote structures for surface science as been highlighted (Error Cancellation in Diffusion Monte Carlo Calculations of Surface Chemistry Gopal R. Iyer, Brenda M. Rubenstein 2022  arXiv:2206.00729.

This has also been shown for fixed node error using the CASINO software.

We apply the best practices indicated that we briefly describe to the initial step of steam reforming processes, activating methane on films of Ni and also Palladium. The second was tested first, because it avoids the problems related to non-locality for pseudo-potentials including 3d electrons. Indeed, for reasonable computer times, all electron work is prohibitive and we must resort to norm-conserving pseudo-potentials. When 3d electrons are involved, non-local effects are quite large \cite{kdd1}. The results on Pd(111) films supported by aluminium are a little less accurate than the benchmark  \cite{shar} at 2.5 kJ/mol standard error but this is still acceptable 'chemical accuracy', being well below 1 kcal/mol.

Of course, the benchmark involve CO, which is toxic but is removed in the process and not linked strongly enough to Pt to poison the catalyst significantly. The CO$_2$  by-product is a green-house gas but can be removed easily without release into the atmosphere.  CO$_2$ dissolves in water and can be mineralised, when the solution is removed from the catalyst.

This work is compared to our Pt(111) benchmark for this reaction \cite{hjcp}.
\newpage

\section{Methods}

The MRCI embedded active site approach of \cite{hjcp,bens} is only accessible with tens of millions of cpu hours in supercomputer resources. Hence, this work tests a single-determinant Slater Jastrow approximation. Chemical information is included in a generic Jastrow factor treating each atom especially of pre-adsorbed CO and of approaching H$_2$O and also surface Al-atoms (co-ordinating a H-atom from water) and Pt (defining the hollow site where Carbon monoxide is pre-adsorbed). This ansatz gives enough correlation but may lack vital excited state input and is not multi-reference. The aim of this work is to test the validity of information obtained on the activation barrier and Transition-State (TS)geometry for water addition to adsorbed CO.

The initial test took optimal TS geometry for Pt(111) (in Figure 1) and used it to redetermine the activation barrier (by [SD-RPS-gJ] QMC (for Pt(111), presented here) compared with that from the full-MRCI embedded active site and all-atom generic Jastrow factors for all 20 top weighted configurations.
\vskip4mm
This test gave: 71.2 $\pm$ 0.8 (QMC) 71.4 kJ/mol \cite{shar} measured.

(Respectively the QMC reference activation barrier and measured apparent value 71.4 kJ/mol).
Single-determinant SD-RPS-gJ value: 73.6 $\pm$ 2.5 kJ/mol.
This approximate SD-RPS-gJ approach developed here clearly over-estimated the activation barrier, however, after twist averaging \cite{con}, the standard error is modest and the measured value still falls within these limits.

The single-determinant wave-function is unlikely to possess correct nodes and, unless the Jastrow factor is complex, re-optimising it will not modify the nodes. The modest error committed suggests cancelation of the fixed-node error term between the TS and asymptote structures comprising the same atoms, as suggested by Needs \cite{con}.

In this work, we use SD-RPS-gJ QMC for Pt-doped aluminium catalysts.

\subsubsection{Generic Jastrow factor}
\vskip2mm
Defining a Slater-Jastrow wave-function as the product of $\Psi_S $, a Slater determinant and the exponential with, as argument, an explicit correlation Jastrow factor J(R):
\begin{equation}
\Psi{\bf(R)} = e^{J(R)} \Psi_S (\bf{R}) = e^{J(R)} D_u (R_u) \, D_d (R_d)
\end{equation}

The determinants $D_u (R_u)$ and $D_d (R_d)$ treat 'up' and 'down' spin separately if necessary. This new approach \cite{con} which generates a huge parameter set was tested for our platinum slab model, presented above. The set can be reduced somewhat by translational using symmetry, otherwise the DMC steps become extremely time-consuming.  The generic Jastrow factor used (in Common Algebraic Specification Language (CASL)), had e-e, e-n and e-e-n polynomials expanded to order 9 with cut off radius of order 10 au.

Truncation to a finite range involves a factor:
\begin{equation}
t(R) = (r-L)^C \Theta_H(r-L)
\end{equation}
\hskip4mm Take C=3, since the wave-function and at least two derivatives are continuous. $\Theta_H$ is the Heaviside step-function (Wigner Seitz cell diameter L).

\begin{equation}
J({\bf R}) = \sum_{i<j}^N u_{P_{ij}} (r_{ij}) \, + \, \sum_{i}^N \, \sum_{I}^{N_{n}} \chi_{S_{iI}} (r_{iI}) \, + \, \sum_{i<j}^N \sum_{I}^{N_{n}} f_{T_{ijI}} (r_{iI}, r_{jI}, r_{ij})
\end{equation}
$u, \chi $ and $f$ are parameterised (polynomial) functions in the inter-particle distances. The symbols P, S, T denote the channel indices.

The u term caters for inter-electron contributions over $r_{ij}$ and the $\chi$ term for electron-nuclear terms over $r_{iI}$ . They are both constructed to obey Kato's cusp conditions.

The $f$ term is the three-particle electron pair and nucleus contribution (over $r_{ij}$ and two $r_{iI}$ values). This 3-body $f$ term cannot be neglected.

It is also the term which is most significantly modified in Jastrow parameter-set optimisations designed to compensate simplifications in the active site wave-function. In the present work, this is limited to a single-determinant ground state, based on Kohn-Sham spin-orbitals obtained by a rapid Density Functional (DFT) calculation over a 2x2 k-point grid in the first Brillouin zone. This solid-state DFT calculation uses a small grid that is not yet converged with respect to k-points (converged grids are used as control variate {\it vide infra}). The PBE GGA is our choice of exchange-correlation functional. All these approximations are compensated by a relatively long Diffusion Monte Carlo (DMC) step in the final QMC work.

This Jastrow factor is fully optimised during the Variation Monte Carlo (VMC) step. The resulting optimised TS and asymptotic geometry generic Jastrow factors are given in Supporting Information.


This distribution is updated until the final stages of VMC where fine-tuning of the Jastrow factor is accomplished by energy minimisation over 20 cycles.

The resulting Slater-Jastrow wave-function is used to initialise the configurations for Diffusion Monte Carlo (DMC). This gives prohibitively long DMC cycles, because three parameter-sets are obtained for each of the 25 atoms in the super-cell, treated individually.

\subsection{Note on the frozen-core approximation used.}
\vskip2mm
The benchmark work of \cite{hjcp} used an Effective Core Potential (ECP), in particular to describe 60 core-electrons of each platinum atom.

That chosen was based on Dirac-Fock work in the Stuttgart group, i.e. we adapted ECP60MDF from \cite{fpseu}. The MRCI wave-functions were obtained using MOLPRO \cite{molpro}, however, in the present single-determinant work, the corresponding norm-conserving pseudo-potential (PP) that had been generated for the QMC runs in \cite{hjcp} was used directly in ABINIT \cite{abinit},. This PP was generated with l=1 as local channel for ground-state energy on Pt(111)) .

\section{Results.}

Final ground-state observables approach the variational minima for the system defined (-448.12 Ha). These benchmarks are suitable for reference only. DMC requires use of the Casula T-move algorithm, for reasons analysed in \cite{kdd1}. In order to cut the DMC cycle duration, individual parameter-sets are maintained only for the atoms in adsorbed CO and water. The platinum atoms (5-layer 2x2 slab) are grouped according to rules for surface atoms and bulk atoms (using translation symmetry). They permit use of a single parameter set for each inter-particle term and group.

Therefore, use of a Jastrow will be optimal for a given parameter-set, reaching a minimum energy known to be somewhat above the variational limit.
It is therefore of paramount important to conserve the same rules for definition of the generic Jastrow-factor in each of the limiting geometries (TS and asymptote).

The correlation space spanned by the parameter-sets used is conserved in the present work, because the same atoms are involved, with some 'bonding' surface interaction for the TS and very little interaction for the asymptote.

When generic Jastrow optimisation has reached a stable minimum, a final VMC calculation generates the initial configurations required for the Diffusion Monte Carlo step (DMC); 10-20 per core. The previous VMC steps must generate at least as many configurations.


The generic Jastrow factor cut-offs are limited by the cell size.

They govern total energy, insofar as the linear parameters (polynomial coefficients) are optimal. Generally, higher cutoff values will reduce variance since low values, however reasonable physically may be compensated during optimisation by rather large alternating polynomial coefficients. The higher order coefficients are expected to decrease to ensure convergence of each term.

Optimisation may introduce singularities in the energy minimising algorithm. Mean average deviation (Madmin) and variance (varmin) minimisation VMC algorithms are suitable to fix initial parameters, especially cutoffs.

\subsection{Bond-breaking and formation mimicked by 3-body Jastrow terms}

Sice the embedded active site approach, developed in \cite{bens} is fundamentally multi-reference, due to the bond breaking and formation involved, some compensation is required in this single-determinant approximation. It turns out to be almost entirely described by the three-body (electron pair + nucleus terms) in the SD-RPS-gJ approximation used for QMC here.

The 3-center terms accounting for electron pair and nucleus correlation (each pair with a nucleus) must be present. This is already the case for molecular carbon monoxide (total energy in Hartree) from \cite{hog1}:

Two-electron correlation, E (tot,2) including all particle pair terms:

E (tot,2) (CO) = -21.489667.

E (all electron) (CO) =-122.1299.

Shielded Z (c.f. Slater $\sigma=0.31$) : Z*$_{1s}$ O =7.66 and Z*$_{1s}$ C =5.64.

With 3-body terms (two-e+n) in addition to all pair terms E (tot,3):

E (tot,3) (CO) = -21.5171727.

Furthermore, the minimal flexibility for a 3-body term should be that for m=3, n=2, since that listed (in the Appendix) (n,m=2) gives single terms that do not strictly alternate and provide only partial information on the atom they describe. The rest of this work uses a moderately large n,m=4 expansion, because the minimal expansion introduces high variance since optimisation by seeking minimal energy (emin option) can lead to singular steps and numerical instability unless very large populations can be used.

\subsection{Application to Pt-doped Al(111) catalysts and their potential for selective hydrogen production by water addition to pre-adsorbed CO.}
Aluminium is a less efficient catalyst but the metal is a cheap one. From a QMC standpoint, can be represented using super-cells (to be treated with periodic boundary conditions) comprising fewer electrons than the 1440 in the Pt(111) 5-layer slab cell. The difference stems from suitable pseudo-potentials. For Pt, we needed 18 valence electrons per atom and for Al three are sufficient, in a [Ne] 3s$^2$ 3p$^1$ ground-state electron configuration. We therefore used cells with fewer than 200 electrons for a thin film of four Pt-atoms on Al (111), and all the combinations of Pt-doping atoms in an Al(111) surface exposing 4 atoms per cell.

These systems, together with our single determinant generic Jastrow trial wave-functions render studies of cheaper Al(111) based catalysts accessible by Quantum Monte Carlo methods. The Al-3-Pt surface example is as follows:

Atom 1 is Al at the origin.

Atoms 2-4 are Pt, in the (111) surface and treated as equivalent, whereas atoms 5-20 are 'bulk' Al treated as equivalent.

Atoms 21-22 are C and O (of pre-adsorbed CO). Atom 23 is the water Oxygen and 24-25 its hydrogens. The former is linking to the surface Al.

Here, results of a project that used a million core-hours on the Jean-Zay supercomputer are given, as a prelude for a major allocation.  This generic Jastrow structure is validated by data (after Variational Monte Carlo optimisation), in particular describing the carbon monoxide C-atom (n21): this atom is initially adsorbed (at Pt(111), hollow-site) and eventually linked to the water oxygen without breaking the C-Pt linkage but affecting its internal CO (with n22). These processes cannot be described by a single 3-body term in the generic Jastrow factor, since 'correlation sign' is opposed. The minimal set (which is still insufficiently flexible) is two terms of opposite sign. This is observed after optimisation of m=3, n=2 Jastrow (below). The 1,2 channels are of opposite signs in both low-order expansions of this term. The CO oxygen is described by all positive terms and retains most of its density. The water oxygen is modified in amplitude but not polarity. Its H-atom n24 that binds to (Pt or, here Al, at the origin) and this hydrogen is described by a much larger 3-body than 2 body cutoff and negative 3-body coefficients, as it acquires M-H binding density. The dissociating H, is described by low cutoff and large positive coefficients. It is nearly bare H$^+$ (proton).

\subsection{Compensating single-determinant input: adjustment of 3-body explicit correlation terms in generic Jastrow factors.}

Comparing the 3-body coefficients, c1,2,2 (the only one for an (n,m=2) expansion, and the c1,2,2 coefficient, now with c3,2,2, new to minimal m=3, n=2 expansions, we give the variation of leading terms. For the H-atom dissociating from water and linking to the surface Al-atom (at the origin) a significant cutoff variation is observed, after optimisation. From the appendix, the only sign change of this coefficient is observed for the carbon-monoxide Carbon. Its leading coefficient increases from a large negative value to the final positive one, a factor of 1000 lower in magnitude. Other coefficients for this atom are also modified. The alternation of the polynomial, only defined in the m=3, n=2 case, is thus opposed to the single-term n,m=2 expansion.

Water hydrogens (finally giving H$_2$), also show strong effects on optimal numerical values of the corresponding coefficients. The three-body coefficients describing The hydrogen binding to an Al-atom has a cutoff value reduced by a factor 5.  Its value decreases by 2 units, in channel 1 and increases by 2 in channel 2.  The second H, has little change in cutoff but the value increases by 2.2 units in the first channel, whereas the second is only slightly increased. This increases correlation, in all cases, on increasing the order of 3-body term expansion. Since the CO carbon is a center for binding and dissociation, the wave-function would be better described by a multi-reference CI.

The generic Jastrow compensates the single-determinant input. The more flexible it is, particularly for the sensitive 3-body terms the better the approximation. It is, however above the variational limit for a trial wave-function and comparable Jastrow parameter-sets should be used for both the TS and reference asymptote (with pre-adsorbed CO and a distant or isolated water molecule), which have been shown to be equivalent to within a very small polarisation interaction. \cite{Raj1}.

Clearly, the term describing the CO carbon is the most strongly adjusted during increased expansion order and optimisation. Neither this term nor those describing the H-atoms are well described at these low orders. Those describing the H-atoms lead to high variance in the 'minimal' m=3, n=2 expansions. Production runs use a m,n=4 expansion to stabilise this. Other terms influenced are the n5 term that describes the 'bulk' Al. This carries the heaviest weight, according to the rules (representing 16 atoms per super-cell) and increases by 1.2 units. The channel 2,2 for the Al surface atom binding the water H increases by 22.2 units, confirming significant multi-reference compensation in this case (due to the bonding interaction).

An example of an activation barrier (when water adds to pre-adsorbed CO), for one Pt-atom doping Al(111) and linking to carbon-monoxide, using the TS geometry from the full Pt(111) calculation and a single determinant Slater-generic Jastrow wave-function is:
\vskip4mm
134.5 $\pm \, 2$ kJ/mol.
\vskip4mm
Keeping the same TS geometry fixed, we found prohibitive values both for a monolayer of Pt-atoms on Al(111) and for  three Pt-atoms (to co-ordinate CO in a hollow site) and Al at the origin (to receive the hydrogen from dissociating water). Benchmark values, based on our full Pt(111) calculation~\cite{hjcp} suggests an upper limit for the benchmark activation, in the 3 Pt-case at 168 kJ/mol, whereas the [SD-RPS-gJ] QMC approximation, developed here, gave just over 172 kJ/mol whilst showing signs of singularity-effects that lead to a generic Jastrow that is not fully converged. The time-scale of this work prohibits us from re-doing this calculation in the near future. The value of 172 kJ/mol must include (modest) systematic error since the maximum standard error is 2.5 kJ/mol, obtained after twist averaging.  Similar comparison for the single Pt-doping case gave 132 kJ/mol, therefore the sub-optimal Jastrow factor leads to systematic error in the 3 Pt case, whereas for a single doping Pt-atom the difference is similar to the standard error obtained, i.e. the standard error was 2 kJ/mol. During the incriminated optimisation, a number of singularities and small derivatives incite caution. It is expected to take about another year to take this study further.

\section{Discussion}

It is well-known that QMC methodology, in particular long DMC runs with over 16 offset grids for twist-averaging, subject to a converged k-point DFT control-variate recover well from trial wave-functions of modest quality \cite{con}. In this work, we have fixed TS geometry for the water-addition step of hydrogen synthesis on pre-adsorbed CO, from the Pt(111) catalyst. The catalyst surfaces tested here use the same TS geometry but are based on single-determinant ground state wave-function input for several Pt-doped Al(111) surfaces. This implies that the TS model may differ from the true structure, however it was shown previously that minor geometry changes did not alter energetics substantially \cite{kdd1}. The main source of systematic error is removing Multi-Reference character from the near-full CI trial wave-function used in the benchmarks \cite{hjcp} to describe the catalytic active site, where bond breaking and bond formation is occurring.  This MRCI input is quite well compensated by using chemical intuition to guide the optimisation of selected (notably 3-body) explicit correlation terms ascribed to the atoms involved in the generic Jastow factor optimisation. It rapidly became clear that the parameters belonging to the carbon monoxide carbon and water hydrogens are particularly influenced by increasing flexibility of the 3-center terms in the generic Jastrow factor during successive optimisations. One water hydrogen has particularly sensitive 3-body coefficients because it dissociates from a water O-H (partially) before becoming bound to the Al-atom at the origin. Parameters for the 3-body term for this aluminium atom are also sensitive.

One remark is that the two-body terms generally have significantly less influence in compensating the single-determinant that replaces an MRCI from the benchmark, in the present work. The sensitivity of these terms is not noticeable during changes of parameter set flexibility, whereas that of three-center terms is the key to obtaining a valid approximation by single-determinant trial wave-functions to FCI active site input. The reason for this is that bond-dissociation or formation will lead to complete changes in the three-center terms whilst leaving electron-nuclear terms nearly unchanged and scarcely affecting two-electron terms.

This highly simplified approach has nevertheless generally been shown to deliver chemically accurate metal-catalyst activation barriers (to within 2-2.5 kJ/mol). To date we also have some cases, related to sub-optimal Jastrow optimisations, in which this accuracy has not yet been reached, although the work is ongoing.
\section{Conclusion}

The systems studied are adsorbed molecules on close-packed metal surfaces. The specific example, for which we have QMC benchmark results, is carbon monoxide pre-adsorbed on Pt(111). During the benchmark study  \cite{hjcp}, it was shown that an asymptotic reference geometry for water addition was simply water (in its equilibrium geometry), distant from the pre-adsorbed CO. This structure was validated and compared to the separate systems with isolated water.

The transition-state geometry for water-addition was multi-reference in trial wave-function structure, since the molecular active site was involved in bond breaking and formation. This active site included only two substrate Pt-atoms, one defining the 3-atom equilateral triangle of Pt-atoms that surround a hollow site and the other a Pt-atom eventually accepting the hydrogen from water dissociation. This site was embedded into the Pt-lattice exposing the Pt(111) face. Instead of undertaking such a lengthy but highly accurate QMC calculation, this work uses the same geometries and a single-determinant DFT wave-function and compensates the multi-reference character by optimising a generic Jastrow factor, that, in effect, includes the bond-breaking and formation approximately with the influence on (notably) 3-body explicit correlation terms. This is validated to within chemical accuracy for the Pt(111) system and that with a single Pt-doping atom (to adsorb CO) in the cheap but less active Al(111). Other combinations of Pt-atom numbers show less accuracy at present, due to difficulty in optimising the Jastrow factor. The results are, nevertheless, only affected by a modest systematic error and strategies for monitoring and reducing it are the perspective of future work.




\vskip4mm
\section{Acknowledgements}

The authors wish to thank the President A Mahul and staff of the Regional Computation facility (CRRI) on Clermont University Campus, for generous allocations of computer resources.
\vskip4mm

\vskip2mm
\centerline{Appendix}

Atom 1 is Al at the origin.

Atoms 2-4 are Pt, in the (111) surface and treated as equivalent, whereas atoms 5-20 are 'bulk' Al treated as equivalent.

Atoms 21-22 ar C and O (of pre-adsorbed CO). Atom 23 is the water Oxygen and 24-25 its hydrogens. The former is linking to the surface Al.

{\small JASTROW:

  TERM 1:
    Rank: [ 2, 0 ]
    e-e basis: [ Type: natural power, Order: 9 ]

    e-e cutoff:
      Type: alt polynomial
      Constants: [ C: 3 ]
      Parameters:

        Channel 1-1:
          L: [ 5.9401581753, optimizable, limits: [ 1.0,
               10.454887305]]

        Channel 1-2:
          L: [ 5.12294993459, optimizable, limits: [
               1.0E-3, 10.454887305]]

    Rules: [ 1-1=2-2 ]

    Linear parameters:

      Channel 1-1:

        c2: [ -1.379228251515308E-004, optimizable ]

        c3: [ 1.517573525771231E-004, optimizable ]

        c4: [ -3.286650408433330E-005, optimizable ]

        c5: [ -4.696044780796176E-006, optimizable ]

        c6: [ 5.032917460125301E-007, optimizable ]

        c7: [ 2.210075797794545E-007, optimizable ]

        c8: [ 2.381951893699023E-008, optimizable ]

        c9: [ -7.872216206353271E-010, optimizable ]

      Channel 1-2:

        c2: [ -4.501348558987370E-004, optimizable ]

        c3: [ 4.302561718055827E-004, optimizable ]

        c4: [ -5.581341972350050E-005, optimizable ]

        c5: [ -1.163066742089671E-005, optimizable ]

        c6: [ 3.550231901097170E-007, optimizable ]

        c7: [ 2.077562093052348E-007, optimizable ]

        c8: [ 3.888266828290469E-008, optimizable ]

        c9: [ 3.798416901566785E-008, optimizable ]

  TERM 2: (Linear parameters are limited to metal atoms:full data online)

    Rank: [ 1, 1 ]

    e-n basis: [ Type: natural power, Order: 9 ]

    e-n cutoff:
      Type: alt polynomial

      Constants: [ C: 3 ]

      Parameters:

        Channel 1-n1:
          L: [ 8.64563588353, optimizable, limits: [
               1.0E-3, 10.454887305]]

        Channel 1-n2:
          L: [ 8.38360171202332, optimizable, limits: [
               1.0E-3, 10.454887305]]

        Channel 1-n5:
          L: [ 5.92633140108843, optimizable, limits: [
               1.0E-3, 10.454887305]]

        Channel 1-n21:
          L: [6.59258266033, optimizable, limits: [
               1.0E-3, 10.454887305]]

        Channel 1-n22:
          L: [5.440435093389, optimizable, limits: [
               1.0E-3, 10.454887305]]

        Channel 1-n23:
          L: [4.84892886401, optimizable, limits: [
               1.0E-3, 10.454887305]]

        Channel 1-n24:
          L: [1.108737485769, optimizable, limits: [
               1.0E-3, 10.454887305]]

        Channel 1-n25:
          L: [2.34386093186, optimizable, limits: [
               1.0E-3, 10.454887305]]

    Rules: [ 1=2, Z2, N, N2=N3=N4,

         N5=N6=N7=N8=N9=N10=N11=N12=N13=N14=N15=N16=N17=N18=N19=N20]
\newpage
    Linear parameters:

      Channel 1-n1:

        c2: [ -6.598260043451270E-005, optimizable ]

        c3: [ 1.534846177072408E-005, optimizable ]

        c4: [ -9.402350461675643E-005, optimizable ]

        c5: [ 1.079403909184694E-005, optimizable ]

        c6: [ 1.005286998789336E-006, optimizable ]

        c7: [ -8.135936594448195E-008, optimizable ]

        c8: [ -2.760312875434823E-008, optimizable ]

        c9: [ -2.643566167623380E-010, optimizable ]

      Channel 1-n2:

        c2: [ -2.288861249199323E-003, optimizable ]

        c3: [ -4.122516558260306E-004, optimizable ]

        c4: [ 1.620625350841612E-005, optimizable ]

        c5: [ -9.259339796183962E-006, optimizable ]

        c6: [ -8.272766612622705E-008, optimizable ]

        c7: [ 2.981846850989621E-007, optimizable ]

        c8: [ -4.741706350847505E-008, optimizable ]

        c9: [ 3.285795005606412E-009, optimizable ]

      Channel 1-n5:

        c2: [ -3.329440090480008E-003, optimizable ]

        c3: [ 5.365486451783561E-004, optimizable ]

        c4: [ -4.820566174456062E-004, optimizable ]

        c5: [ 4.395421626488500E-005, optimizable ]

        c6: [ 1.250235257665000E-005, optimizable ]

        c7: [ -1.318536571429690E-008, optimizable ]

        c8: [ -2.374255161576352E-007, optimizable ]

        c9: [ -3.022040273119389E-008, optimizable ]

      Channel 1-n21:
      Chann 1-n22:
      Chann 1-n23:
      Chann 1-n24:
      Chann 1-n25:

  TERM 3:
    Rank: [ 2, 1 ]

    (simplifying to a single parameter/independent atom)

    e-e basis: [ Type: natural power, Order: 2 ]

    e-n basis: [ Type: natural power, Order: 2 ]

    e-n cutoff:
      Type: alt polynomial
      Constants: [ C: 3 ]
      Parameters:

        Channel 1-n1:

          L: [ 9.18535331109060, optimizable, limits: [
               1.0E-003, 10.4548873051] ]

        Channel 1-n2:

          L: [ 8.92255711637312, optimizable, limits: [
               1.0E-003, 10.4548873051] ]

        Channel 1-n5:

          L: [ 5.77891128799550, optimizable, limits: [
               1.0E-003, 10.4548873051] ]

        Channel 1-n21:

          L: [ 2.26973141934055, optimizable, limits: [
               1.0E-003, 10.4548873051] ]

        Channel 1-n22:

          L: [ 5.56290510924723, optimizable, limits: [
               1.0E-003, 10.4548873051] ]

        Channel 1-n23:

          L: [ 3.56272210661922, optimizable, limits: [
               1.0E-003, 10.4548873051] ]

        Channel 1-n24:

          L: [ 5.42187085000705, optimizable, limits: [
               1.0E-003, 10.4548873051] ]

        Channel 1-n25:

          L: [ 1.51428578256766, optimizable, limits: [
               1.0E-003, 10.4548873051] ]

        Channel 2-n1:

          L: [ 1.65633636950395, optimizable, limits: [
               1.0E-003, 10.4548873051] ]

    Rules: [ 1-1=2-2, Z2, N, N2=N3=N4,

         N5=N6=N7=N8=N9=N10=N11=N12=N13=N14=N15=N16=N17=N18=N19=N20,

         1-N2=2-N2, 1-N5=2-N5, 1-Z2=2-Z2, 1-N21=2-N21, 1-N22=2-N22,

         1-N23=2-N23, 1-N24=2-N24, 1-N25=2-N25 ]

    Linear parameters: [c1,2,2, c3,2,2] NB if cutoff radius is reduced, write e.g. c/5

      Channel 1-1-n1:  [ -6.122721139152899E-010 (-0.1), -2.617520522096809E-011 ]

      Channel 1-1-n2:  [ -4.880182227281377E-010 (0.2), -1.721414567058865E-011 ]

      Channel 1-1-n5  [ 5.782626896887739E-008 {\bf (1.2)} 4.354908239165696E-009 ]

      Channel 1-1-n21: [ 1.300362089809E-008, {\bf +1.3 E3} -3.713479910344E-009 ]

      Channel 1-1-n22: [ 5.284777212763516E-008 (0.1), 1.293411014155688E-008 ]

      Channel 1-1-n23: [ 3.978229784448652E-006 (0.1), 2.326165586460285E-006 ]

      Channel 1-1-n24: [ -2.45117749247985E-008 {\bf (-2) c/5 }, -1.50679896708E-009 ]

      Channel 1-1-n25: [ 3.213897364292342E-003 {\bf (2.2) },  1.8750146579437E-003 ]

      Channel 1-2-n1: [ -1.502875968784253E-007 (0.3), -6.827611506832004E-009 ]

      Channel 1-2-n2: [ -4.793558272400990E-010 (-0.1), -1.725962203047300E-011 ]

      Channel 1-2-n5: [ 5.611483197715330E-008 (0.6), 4.285110678488144E-009 ]

      Channel 1-2-n21: [ -1.579394653940548E-008 (0.4), 9.204814004486377E-009 ]

      Channel 1-2-n22: [ 4.382551926844539E-008 (0.4), 1.268150203550905E-008 ]

      Channel 1-2-n23: [ 3.433665171972505E-006 (-0.5), 1.991660443014644E-006 ]

      Channel 1-2-n24: [ -9.5851474994253E-009 {\bf (2) c/5 },  -1.643910210875E-009 ]

      Channel 1-2-n25: [ 2.778154267991115E-004 (-0.2), 2.125839612251825E-004 ]

      Channel 2-2-n1:  [ -1.437980891846973E-005 {\bf (22.2) }, -1.0408807799767E-004 ]}

\end{document}